\documentclass[]{JHEP3}
\usepackage{epsfig,multicol,bbm}

\title{On Wald entropy of black holes: logarithmic corrections and trace anomaly}
\author{R Aros, D E D\'{\i}az and A Montecinos\\Universidad Andres Bello,
Departamento de Ciencias Fisicas,
Republica 220, Santiago, Chile\\
	E-mail: \email{raros,danilodiaz,alejandramontecinos@unab.cl}}

\abstract{Quantum effects due to conformal matter in a black hole background result in universal logarithmic corrections
to black-hole entropy. The universality resides in the connection of the log term coefficient with those of type-A and
type-B Weyl anomalies, regularization-scheme independent quantities.
We presently study the case of extremal black holes within Wald's Noether charge formalism. In the conformal class of
flat metrics, we are again able to unveil the log term in the entropy from the horizon value of the solution to the
Q-curvature uniformization problem.
Beyond conformally flat backgrounds, type-B Weyl anomaly becomes an obstruction to considering flat space as the fiducial
metric and the search for a metric of constant Q-curvature remains open.
Notwithstanding, by a uniform scaling argument we show that the results based on heat kernel and euclidean computations
(namely entropy function and conical defect) can also be derived as Wald entropy, that is, as Noether charge of the
integrated anomaly or conformal index.
We finally comment on the relation with entanglement entropy.}

\begin{document}


\section{Introduction}
Black holes in Einstein gravity are endowed with entropy given by the celebrated Bekenstein-Hawking (BH)\emph{area law}
\cite{Bekenstein:1973ur, Hawking:1974sw}. Deviations from the area law are naturally expected to occur in more general
classical gravitational theories and also due to quantum effects. This is certainly the case in string theory where
the
effective low-energy theory, in addition to Einstein-Hilbert action, contains higher curvature terms due to finite size
of
strings ($\alpha'$) as well as quantum loop corrections ($g_s$). Here, certain class of extremal black holes do have a dual
microscopic
description and the statistical entropy resulting from the microscopic counting is found to be in remarkable
agreement, for large charges, with the macroscopic counterpart that takes into account higher \emph{local} curvature
invariants
from finite-size and quantum corrections (see, e.g., \cite{Maldacena:1997de,Lopes Cardoso:1998wt,LopesCardoso:1999cv}).
Crucial to this agreement is the fact that the corresponding macroscopic entropy,
which deviates from the area law, is \emph{Wald entropy}, that is, it satisfies the modified form of the First Law
derived by
Wald~\cite{Wald:1993nt} for a wide class of generally covariant actions; in Wald's formalism, the role of the entropy
is played by
the integral of a geometric density, the Noether potential, over a spatial cross-section of the horizon. In this respect, the
`entropy function formalism'~\cite{Sen:2005iz,Sen:2005wa} has been developed as an efficient computational tool that
exploits the
attractor mechanism for extremal black holes; it can be shown that the entropy function is in fact Wald entropy.

Regarding quantum one-loop effects due to conformal matter, the \textit{leading correction} to the black hole (bh) entropy in
the semiclassical limit of large charges seems to be \textit{universally} given by the logarithm of the horizon area
(cf. \cite{Medved:2004eh,Page:2004xp})
 \begin{equation}
\label{quantumcorrectedEntropy}
 S_{bh} = S_{_{BH}}+ const\cdot\ln S_{_{BH}}+\mathcal{O}(1),
\end{equation}
with a coefficient that involves those of the Type-A and Type-B trace anomalies \cite{Deser:1993yx}.
This deep connection between trace anomaly and entropy had been established in the Euclidean formalism after
renormalization of
one-loop UV-divergent contribution of matter fields to the entropy \cite{Fursaev:1994te,Mann:1997hm}, in the
tunneling and exact
differential formalisms~\cite{Banerjee:2008fz,Banerjee:2009tz}, in the thermodynamics of certain black hole solutions
stemming from
an anomalous energy-momentum tensor~\cite{Cai:2009ua} and, more recently, via the
`quantum entropy function'~\cite{Sen:2008vm}-\cite{Bhattacharyya:2012wz}.
Despite the many different approaches, there was little hope for a direct derivation based on Wald's formalism mainly due
to the unavailability of the one-loop
effective action, which in general contains nonlocal terms.
Notwithstanding, for the conformally flat class of black hole backgrounds we were able to establish the connection
between Type-A
trace anomaly  and the logarithmic correction to the black hole entropy via Wald's Noether charge
formalism~\cite{Aros:2010jb}. The rationale behind our approach was to focus on the
`anomaly-induced effective action' ($^AS_{anom}$), which suffices to correctly
reproduce the anomalous trace of the energy-momentum tensor, and to render it local by the introduction of an auxiliary
field $\phi$
to finally use Wald's formula to read off the contribution to the black hole entropy as Noether charge.

The purpose of the present note is to further elaborate on the Wald approach to log-corrections in two ways.
We first study
the extremal black hole case within the conformally flat class; secondly, we consider the general case where
Type-B trace anomaly comes into play. We are able to obtain the log term, as already done in non-extremal cases, from
the solution the uniformization problem for Q-curvature. Beyond conformal flatness, type-B Weyl anomaly enters the
game and introduces ambiguities in the definition of the Q-curvature; it also becomes an obstruction for flat metric
to be considered as the `fiducial' one and the search for a metric of constant Q-curvature remains open.
Notwithstanding, by a uniform scaling argument we show that the results based on heat kernel and euclidean
computations (namely entropy function and conical defect) can also be derived as Wald entropy, that is, as Noether
charge of the integrated anomaly or conformal index. We finally comment on the relation with entanglement entropy.

\section{Noether charge of the anomaly induced action}

The Type-A trace anomaly in four dimensions is given by the term involving the Euler density in the expectation value of
the trace of the energy-momentum tensor
\begin{equation}
\left\langle T \right\rangle\,= -\frac{a}{16\pi^2}\,E_4~,
\end{equation}
with
\begin{equation}
E_{4}=Riem^2-4Ric^2+R^2~.
\end{equation}
The anomaly induced effective action is a conformal primitive of the trace anomaly that generalizes Polyakov's original computation
in two dimensions~\cite{Polyakov:1981rd}. In analogy with Liouville's local form of Polyakov's action, an auxiliary
field can be introduced to cast the action in local form (see, \textit{e.g.} \cite{Balbinot:1999vg,Mottola:2006ew})
\begin{equation}\label{EffectiveAction}
^AS_{anom} = -\frac{a}{32\pi^2}\int dx^4 \sqrt{-g}\hspace{2ex} \{-\phi\,\Delta_4\,\phi \; +\; Q\,\phi\}~.
\end{equation}
This four dimensional analog involves the $Q-$curvature \cite{BransonFirstQCurvatureApp1991} \footnote{The original
$Q$-curvature
\cite{BransonFirstQCurvatureApp1991} differs by a $Weyl^2-$term and an overall factor of 4 from the one used here,
which is also
the one in~\cite{Riegert:1984kt}, but both have the same linear transformation law under Weyl rescaling of the metric.}
\begin{equation}
Q  = E_{4} - \frac{2}{3} \Delta R~,
\end{equation}
and the Paneitz's operator \cite{Paneitz-2008-4} \footnote{This quartic operator actually first appeared in the physics literature in~\cite{Fradkin:1981jc,Fradkin:1982xc}.}
\begin{equation}
\Delta_4=\Delta^{\,2}+2\,\nabla_\mu \left( R^{\mu\nu}-\frac{1}{3}g ^{\mu\nu}R\right)\nabla_\nu~.
\end{equation}
To correctly reproduce the anomaly from the metric variation of this action, the auxiliary field $\phi$ must satisfy
\begin{equation}\label{EffectiveEquationOfPhi}
    \Delta_4 \phi = \frac{1}{2} \left(E_{4}-\frac{2}{3} \Delta R \right)~,
\end{equation}
that is, $\phi$ must solve the uniformization problem for the Q-curvature with the fiducial metric (flat one) having
vanishing Q-curvature.

The corresponding Noether charge can be computed and yields a leading correction to BH entropy
\begin{equation}\label{ADM-prime}
S_{bh}=\frac{\mathcal{A}_{_\mathcal{H}}}{4} - a\cdot\chi_{_\mathcal{H}}\cdot \phi_{_\mathcal{H}} + \ldots~.
\end{equation}
Here $\chi_{\mathcal{H}}$ is the (2 dimensional) Euler characteristic of the horizon,
$\phi_{_\mathcal{H}} = \phi(r_{+})$
is the value at the horizon of the auxiliary field and the ellipsis stands for terms involving only derivatives of
$\phi$
at the horizon. The derivatives of $\phi$, and its powers, translate into inverse powers of the area, so that they
do not talk
to the logarithmic correction\footnote{The vanishing contribution form terms such as $(\Delta\phi)^2$ was noticed
previously in \cite{Jacobson:1993vj}.} and are absorbed in the $\mathcal{O}(1)$ term.

\section{Extremal black holes: near-horizon geometry}

We start by considering the near-horizon geometry of a black hole in the extremal limit. It is well known that
this geometry factorizes in an $AdS_2$ times a transverse compact space
\cite{Zaslavsky:1997ha,Mann:1997hm} but the limiting procedure has its own subtleties.
Let us briefly examine this limit in the following simple Euclidean spherically symmetric situation, described by
the metric
\begin{equation} \label{spherically}
ds^2=f(r)d\tau^2+f(r)^{-1}dr^2+r^2d\Sigma
\end{equation}
where $d\Sigma$ is the metric of the compact transverse part. If we expand around the outer horizon $r_+$
\begin{eqnarray}
f(r)&=&a(r-r_+)+b(r-r_+)^2+O((r-r_+)^3)\\
&=&(r-r_+)\left(r-r_++\frac{a}{b}\right)b+O((r-r_+)^3)~,
\end{eqnarray}
the extremal limit is then achieved by sending $a\rightarrow0$ (i.e. $r_-\rightarrow r_+$) and rescaling
the Euclidean time coordinate $\tau$, where $\tau\sim\tau+4\pi/a$. In suitable coordinates $x,\phi$
\begin{eqnarray}
r-r_+&=&\frac{a}{b}\sinh^2\frac{x}{2}~,\\
\frac{a\tau}{2}&=&\theta, \qquad \theta\sim\theta+2\pi~,
\end{eqnarray}
a two dimensional hyperbolic $H^2$ factor shows up
\begin{equation} \label{hdos}
ds^2=\frac{1}{b}\left( dx^2 + \sinh^2x\,d\theta^2\right) + r_+^2 d\Sigma^2~.
\end{equation}
The horizon is mapped to the center $x=0$ of $H^2$ and the conformal boundary to $x\rightarrow\infty$.
\paragraph{Extremal Reissner-Nordstr\"om:}

As illustration let us examine the conformally flat near-horizon geometry $H^2\times S^2$ of the extremal
Reissner-Nordstr\"om
\begin{equation} \label{RNE}
ds^2=r_+^2\left( dx^2 + \sinh^2x\,d\theta^2 + d\Omega^2\right)~.
\end{equation}
In order to solve the uniformization problem for the Q-curvature, we choose Fefferman-Graham coordinates and set
$r_+=1$ for convenience,
\begin{equation}
    \frac{\rho}{2}=\mbox{e}^{-x}\,;\qquad x=0\mapsto \rho=2 \,;\qquad x\rightarrow\infty\mapsto \rho\rightarrow 0;
\end{equation}
so that the $H^2$ metric becomes
\begin{equation}\label{FFG}
   ds^2_{H}=\frac{d\rho^2+(1-\rho^2/4)^2d\theta^2}{\rho^2}\,.
\end{equation}
The Paneitz operator gets factorized in terms of the Laplacian as $\Delta_4=\Delta^2-2\Delta$ and the solution,
which is required to be regular at the horizon, has the following logarithmic term
\begin{equation}
    \phi(\rho)\sim 2\ln\frac{(\rho+2)^2}{\rho}~.
\end{equation}
Now we  read off the logarithmic dependence on $r_+$, $\phi_\mathcal{H}\sim 2\ln r_+\sim \ln A_\mathcal{H}$,
and the logarithmic correction to the entropy in (\ref{ADM-prime})
\begin{equation}\label{ERN-NH}
    \triangle S_{bh}=-2\,a\ln A_\mathcal{H}~.
\end{equation}

\section{Beyond conformal flatness: type-B Weyl anomaly}

When the background is not conformally flat, the type B Weyl anomaly comes into play and things are more complicated
\begin{equation}
\left\langle T \right\rangle\,= -\frac{a}{16\pi^2}\,E_4+\frac{c}{16\pi^2}\,W^2~.
\end{equation}
The anomaly induced effective action becomes ambiguous, essentially due to the ambiguity in the Q-curvature.
One is free to shift any Q-curvature by the locally conformal invariant Weyl-squared term while preserving the
linear transformation law under Weyl rescaling of the metric $g\rightarrow \hat{g}=e^{2w}g$
\begin{equation}
    \sqrt{\hat{g}}\,\hat Q= \sqrt{g}\left( Q+\Delta_4\, w\right)\qquad \mbox{and}\qquad
    \sqrt{\hat{g}}\,\hat{W}^2=\sqrt{{g}}\,{W}^2~.
\end{equation}
This ambiguity carries over into the uniformization problem and we have, a priori, no preferred choice for the Q-curvature.\\

Now, as long as one is interested in the dependence on an overall scale factor, things are easier. Consider a
uniform scaling of the metric  $g\rightarrow \hat{g}=\Lambda^2 g$; the variation of the one-loop effective action
(log of the functional determinant of the kinetic operator $A$ of the free conformal field) is given by the associated zeta
function at zero (see, e.g.~\cite{Hawking:1976ja})
\begin{equation}
    \hat{S}_{eff}-S_{eff}=-\frac{1}{2}\log\frac{\det\hat A}{\det A}=\zeta_A(0)\,\log\Lambda~.
\end{equation}
Modulo zero mode contributions, this is essentially the integrated trace anomaly or conformal index
\begin{equation}
    \zeta_A(0)=\int\sqrt g\, d^4x\langle T\rangle=\frac{1}{16\pi^2}\int\sqrt g\, d^4x(-aE_4+cW^2)~.
\end{equation}
We simply apply Wald's prescription to translate the dependence on the overall scale $\Lambda$ from the
effective action into the black hole entropy. That is, we compute the Noether charge of the above local quantity.

It is convenient to split the Weyl-square term as follows
\begin{equation}
    W^2=E_4+2\mbox{Ric}^2-\frac{2}{3}\mbox{R}^2~,
\end{equation}
and compute the Noether charge of
\begin{equation}
    \Delta S_{eff}=-\frac{\log\Lambda}{16\pi^2}\int\sqrt g\,d^4x\left((a-c)\,E_4+\frac{2c}{3}\,\mbox{R}^2-2c\,
    \mbox{Ric}^2\right)~.
\end{equation}
The contribution of the first term in the RHS gives, upon integration, the Euler characteristic of
the horizon ($\chi_{_{\mathcal{H}}}$); the second, two times the Ricci scalar (R) of the four dimensional
space; and the third, the Ricci scalar of the four dimensional space along the transverse directions to the
horizon \cite{Jacobson:1993vj,Aros:2001gz}
\begin{equation}
    \Delta S_{bh}=-\frac{\log\Lambda}{16\pi^2}\,4\pi\int_{\Sigma} \sqrt h\,d^2x\left(2(a-c)\,R_{_{\Sigma}}+
    \frac{4c}{3}\,\mbox{R}-2c\,\mbox{R}_{aa}\right)~.
\end{equation}
where R$_{aa}$ is the projection of the Ricci scalar perpendicular to the horizon.

In all, we end up with the following general relation between the correction to black hole entropy,
logarithmic in the overall scale, and the Type-A and Type-B anomaly coefficients
\begin{equation}
    \Delta S_{bh}=\log\Lambda^2\left((c-a)\,\chi_{_{\mathcal{H}}}-\frac{2c}{3}\,\hat{\Sigma}\,
    \mbox{R}^{^{\mathcal{H}}}+c\,\hat{\Sigma}   \,\mbox{R}_{aa}^{^{\mathcal{H}}}\right)~.
\end{equation}
where $\hat{\Sigma}$ stands for the horizon area over $4\pi$ and $R^{^{\mathcal{H}}}$ and $R_{aa}^{^{\mathcal{H}}}$
are the average values on the horizon of the corresponding quantities. In what follows, let us examine some
illustrative examples.

\paragraph{Extremal Reissner-Nordstr\"om:}
This is just a crosscheck, the near-horizon geometry $H^2\times S^2$ is conformally flat and therefore Type-B
anomaly plays no role.
The necessary inputs are
\begin{equation}
    \chi_{_{\mathcal{H}}}=2, \qquad \hat{\Sigma}=1, \qquad \mbox{R}^{^{\mathcal{H}}}=0,\qquad
    \mbox{R}_{aa}^{^{\mathcal{H}}}=-2~,
\end{equation}
so that the log-correction is given by
\begin{equation}
    \Delta S_{bh}=-4\,a\log\Lambda~,
\end{equation}
in agreement with what we obtained before (\ref{ERN-NH}), noting that the area scales as $\Lambda^2$, and with
the results via conical defect~\cite{Mann:1997hm}.
It is worth to notice that the answer remains the same even if one considers the full geometry of the extremal Reissner-Nordstr\"om and not
just the near-horizon one, very reminiscent of an underlying attractor mechanism.
\paragraph{Extremal topological black hole:}
We consider a family of extremal topological black holes~\cite{Aros:2000ij}.
The near-horizon geometries of these black holes have the form $H^2\times T^2$ and $H^2\times H^2/\Gamma$,
the transverse compact space is a space of zero or constant negative curvature.\\

\emph{Torus:} Starting with the torus, the inputs to the log-correction expression are
\begin{equation}
    \chi_{_{\mathcal{H}}}=0, \qquad \hat{\Sigma}=1, \qquad \mbox{R}^{^{\mathcal{H}}}=-2,\qquad
    \mbox{R}_{aa}^{^{\mathcal{H}}}=-2~,
\end{equation}
resulting in
\begin{equation}
    \Delta S_{bh}=-\frac{4\,c}{3}\,\log\Lambda~.
\end{equation}
This result is in conformity with~\cite{Mann:1997hm} when one modifies the heat kernel computation to account for the
conformal coupling of the scalar field.\\

\emph{Hyperbolic surface:} For the case of a transverse hyperbolic surface of genus $g$
\begin{equation}
    \chi_{_{\mathcal{H}}}=2-2g, \qquad \hat{\Sigma}=g-1, \qquad \mbox{R}^{^{\mathcal{H}}}=-4,\qquad
    \mbox{R}_{aa}^{^{\mathcal{H}}}=-2~,
\end{equation}
resulting in
\begin{equation}
    \Delta S_{bh}=4\,(a-\frac{2}{3}\,c)\,(g-1)\,\log\Lambda~,
\end{equation}
in agreement with~\cite{Mann:1997hm} adapted to the conformal scalar field.

\paragraph{Schwarzschild:} This case is also amenable to treat, since there is only one dimensionful parameter,
namely the radius of the horizon,
\begin{equation}
    \chi_{_{\mathcal{H}}}=2, \qquad \hat{\Sigma}=1, \qquad \mbox{R}^{^{\mathcal{H}}}=0,\qquad
    \mbox{R}_{aa}^{^{\mathcal{H}}}=0~,
\end{equation}
resulting in
\begin{equation}
    \Delta S_{bh}=4\,(c-a)\,\log\Lambda~,
\end{equation}
which is easily compared with \cite{Fursaev:1994te}~\footnote{Let us take this opportunity to
correct a mistake in our previous work~\cite{Aros:2010jb}. The uniformization problem in Schwarzschild geometry cannot be
treated with the flat metric as fiducial. There is a minus sign mismatch with the correct value obtained here by scaling
argument.}.

\paragraph{Reissner-Nordstr\"om:}
The non-extremal Reissner-Nordstr\"om, containing Schwarzschild and the extremal RN as particular cases, can
also be worked out
\begin{equation}
    \chi_{_{\mathcal{H}}}=2, \qquad \hat{\Sigma}=1, \qquad \mbox{R}^{^{\mathcal{H}}}=0,
    \qquad \mbox{R}_{aa}^{^{\mathcal{H}}}=-\frac{2r_-}{r_+}~.
\end{equation}
The log term is then given by
\begin{equation}
    \Delta S_{bh}=4\,\left(c-c\,\frac{r_-}{r_+}-a\right)\log\Lambda~.
\end{equation}
This is in concordance with~\cite{Solodukhin:1994st}, for the scalar field ($a=1/360,c=1/120$)
\begin{equation}
    \Delta S_{bh}=\frac{2r_+-3r_-}{90r_+}\,\log\Lambda~.
\end{equation}

\section{Entanglement entropy}
It is a general belief that the proper interpretation of entanglement entropy in a black hole background is that of
quantum correction to black hole entropy, with the event horizon playing the role of entangling surface.
We notice that recent computations of entanglement entropy~\cite{Ryu:2006ef} account for the connection between the
logarithmic-in-the-cutoff term and trace anomaly coefficients. Also by means of a scaling argument, one can track down the
trace anomaly dependence on the entanglement entropy; the computation requires the deformation of the background geometry
by a conical defect. The final results (see also~\cite{Solodukhin:2008dh,Myers:2010tj}) coincide with those derived here
as Wald entropy~\footnote{It is worth mentioning that entanglement entropy for a general entangling surface is far more
complicated in general, depending also on the extrinsic curvature~\cite{Solodukhin:2008dh,Hung:2011xb}.}; in fact, one can
even show \emph{a posteriori} that they are precisely given by Wald's formula~\cite{Hung:2011xb}.
The subtle difference with our approach is that we claim the correction to be Wald entropy from the outset.
In all, we get another instance where Wald formula agrees with euclidean and other approaches whithin their domains of
applicability.

\section{Conclusion}\label{ConclusionSec}

The purpose of this short note was to further study the connection between trace anomaly and quantum corrections
to black hole entropy within Wald's Noether charge formalism. The plan was to pursue the universality of
this connection: even if in general it is not possible to know the full one-loop effective action, the part
that is able to correctly reproduce the trace anomaly did suffice to track down the effect of the anomaly.
This piece is given by the anomaly induced effective action which is rendered local by means of an auxiliary field.
In conformally flat backgrounds, where only type-A trace anomaly plays a role, the on-shell condition for this auxiliary
field corresponds to the mathematical problem of finding a Weyl-scaling of the black hole metric to a fiducial one
which in this case is nothing but the flat one where the Q-curvature vanishes. We showed that even in the extremal
limit one is able to find a solution regular at the horizon and read off a term logarithmic in the horizon area.
Beyond conformal flatness some difficulties are encountered, the non-vanishing Weyl tensor becomes an obstruction to
considering flat metric as the fiducial one and the definition of the Q-curvature is plagued by ambiguities that,
at this moment, we have been unable to tackle~\footnote{One could get rid of the fiducial or reference metric, but not of the ambiguities, 
by considering instead a dilaton effective action~\cite{Schwimmer:2010za,Komargodski:2011vj}; however, the price to pay is to have 
a rather cumbersome nonlinear equation of motion for the dilaton field.}.

A more modest progress is achieved by considering the dependence on an
overall scale. The change in the effective action is logarithmic in the scale factor and contains a local conformal
invariant functional, namely, the integrated trace anomaly or conformal index. The effect of this term on the entropy
can then be obtained $\grave{a}\,la$ Wald by computing its contribution to the Noether charge. We find agreement between this
direct
derivation via Wald entropy and the logarithmic terms in the entanglement entropy; the latter being derived by
deformation
of the effective action in the presence of a conical defect.

Finally, let us comment that agreement is also found in the case of extremal black holes with the  results via
`quantum entropy function', at least the dependence of the logarithmic term on the trace
anomaly~\cite{Sen:2008vm}-\cite{Bhattacharyya:2012wz}\footnote{There are further contributions due to detailed analysis of zero
modes, statistical ensemble and other than purely geometric backgrounds.}. At the level of
the classical action, the entropy function is precisely Wald entropy; our present derivation lends
support to considering that the quantum entropy function is also Wald entropy of the 1-loop effective action.

\section*{Acknowledgments}

This work was partially funded through Fondecyt-Chile 11110430, 1131075 and UNAB DI-21-11/R, DI-286-13/R, DI-295-13/R.


\hspace{0.5cm}


\providecommand{\href}[2]{#2}\begingroup\raggedright\endgroup
\end{document}